\documentclass[aps,twocolumn,superscriptaddress]{revtex4}

\usepackage{amsmath,amssymb,bm}
\usepackage{graphicx}
\usepackage{amsfonts}
\usepackage{color}
\usepackage{color}

     \graphicspath{{./figs/}}

\newcommand{\nn}{\nonumber}                                           
\newcommand{\va}[1]{\langle{#1}\rangle}                               

\newcommand{\al}{\alpha}
\newcommand{\be}{\beta}
\newcommand{\ga}{\gamma}

\newcommand{\ro}{\rho}

\newcommand{\Tr}{\mathop{\rm Tr}\nolimits}

\begin{document}
\thispagestyle{empty}
 \date{\today}
  \preprint{\hbox{...}}

\title{Is the exotic $0^{--}$ glueball a pure gluon state ?}

 \author{Alexandr Pimikov}
 \email{pimikov@mail.ru}
 \affiliation{Institute of Modern
    Physics, Chinese Academy of Science, Lanzhou 730000, China}
 \affiliation{Bogoliubov Laboratory of Theoretical Physics, Joint
    Institute for Nuclear Research,\\ Dubna, Moscow Region, 141980
    Russia}

  \author{Hee-Jung Lee}
  \affiliation{Department of Physics Education, Chungbuk National
    University, Cheongju, Chungbuk 361-763, Korea}

  \author{Nikolai Kochelev}
  \email{kochelev@theor.jinr.ru}
  \affiliation{Institute of Modern Physics,
    Chinese Academy of Science, Lanzhou 730000, China}

  \affiliation{Bogoliubov Laboratory of Theoretical Physics, Joint
    Institute for Nuclear Research,\\ Dubna, Moscow Region, 141980
    Russia}
   \affiliation{Department of Physics, Tokyo Institute of Technology, Meguro, Tokyo 152, Japan}

  \author{Pengming Zhang}
  \affiliation{Institute of Modern Physics,
    Chinese Academy of Science, Lanzhou 730000, China}


\begin{abstract}
We present a new calculation of the mass and width of the exotic $0^{--}$ glueball
 in the framework of the  QCD sum rules.  We next construct a new current which couples to a pure  $0^{--}$ gluon state
 and derive consistent and stable sum rules. A previously used current in this approach was shown to be inconsistent.
We obtain  for this state a mass  $M_G=6.3^{+0.8}_{-1.1}$ GeV and an  upper limit for the total width $\Gamma_G\leq 235$~MeV.
These values can be used as an important guide for the experimental  search of this exotic state.
We argue that the mixing of this glueball state with $0^{--}$ tetraquark is very small.
Therefore, the exotic $0^{--}$ glueball can be considered as a pure gluon state.

\end{abstract}
\pacs{12.38.Lg, 12.38.Bx}
\keywords{Glueball, oddball, QCD sum rules, condensates}

\maketitle

The glueballs  carry  very  important information  on the gluonic sector of QCD and their study
is one of the fundamental tasks  in strong interaction physics.
While glueballs are predicted by QCD, there  has been no clear experimental evidence
of their existence and so  they remain  as of yet   a subject
to theoretical and experimental research (see reviews ~\cite{Mathieu:2008me,Ochs:2013gi}).
For this  reason the study of  glueball candidates is included in many  programs of
presently running and future experiments.

One of the main problems of glueball spectroscopy is the
 mixing of the glueballs with ordinary meson states,  which leads to difficulties  in disentangling the glueball components  in  experiments.
In  this connection, the discovery of the  exotic  $0^{--}$ glueball would be extremely useful,
because it does not mix with any  $q\bar q$ states.
it is therefore very important to investigate the properties of this glueball within a QCD based approach.
One of the most successful approaches to study strong interaction spectroscopy is the QCD Sum Rules (SRs) method~\cite{Shifman:1978bx}.

In this Letter, for the first time, a consistent SR for the exotic $0^{--}$ glueball is obtained.
We calculate the Operator Product Expansion (OPE) of the correlator  up to
dimension-8 with a new interpolating current
which couples to this pure gluon state,  and show that there is good stability for the SR.
From this stable SR a prediction for the mass and an upper limit of the total width of this state are found.

The  QCD SR approach~\cite{Shifman:1978bx} for a bound state consists of two parts.
One is the calculation of the OPE of the correlator defined by
\begin{equation}\label{correlator} 
\Pi(Q^2) = i\int\!\! d^4x\, e^{iqx} \va{J(0)J^\dagger(x)}\,
\end{equation}
where the current couples to the gluonic bound state $|G\rangle$ in our case as
\begin{eqnarray}\nn 
\va{0|J|G}=F_G M_G^{N-2}.
\end{eqnarray}
Here $Q^2=-q^2$, $N$ is the dimension of the current $J$,
$F_G$ is the decay constant and $M_G$ is the mass of the state.
To construct the SR we follow for the second part, usually called phenomenological part,  the pioneering work of ref.~\cite{Shifman:1978bx} and
the recent study of the scalar and pseudoscalar glueballs by Forkel~\cite{Forkel:2003mk}.
Putting these pieces together, the corresponding SR for a zero width resonance model of the spectral density, ($\rho\sim \delta(s-M_G^2)+$
continuum), has the following form:
\begin{eqnarray}\label{SR1}
\frac{1}{\pi}\int_0^{s_0}\frac{\text{Im}\Pi_{\text{(OPE)}}(-s)}{s+Q^2}ds
    &=&
    \frac{F_G^2M_G^{2(N-2)}}{M_G^2+Q^2}\,,
\end{eqnarray}
where  $\Pi_{\text{(OPE)}}(-s)$ is the OPE of the correlator,
Eq. (\ref{correlator}),
and  $s_0$ is the continuum threshold.
It is known that $0^{--}$ state can not couple to a three-gluon interpolating current
without derivatives \cite{Jaffe:1985qp}.
In the paper~\cite{Qiao:2014vva} a very specific current with derivatives has been constructed
to obtain the mass of three-gluon exotic glueball.
However, in~\cite{Pimikov:2017xap} it has been demonstrated that this current leads to
the inconsistency of QCD SR.
Here we propose a new gauge invariant current with derivatives which couples to the $0^{--}$ state.
 It has the general form:
\begin{equation}\label{eq:Jmm-general}
    J(x)=\frac 23  g_s^3  
    \epsilon^{ijk}
    \Tr\left(
    (O_i G_{\mu\nu}(x))
    (O_j G_{\nu\ro}(x))
    (O_k G_{\ro\mu}(x))
    \right)\,,
    \end{equation}
where  $G^{a}_{\mu\nu}$ is the field strength tensor,
$\tilde G^{a}_{\mu\nu}\equiv G^{a}_{\al\be}i\epsilon_{\mu\nu\al\be}/2$,
the operators $O_i$ are the products of covariant derivatives $O_i=D_{\alpha_1}\cdots D_{\alpha_{n}}$.
The lowest dimensional current in this form, that  has  nonzero LO perturbative contribution to the SR corresponds to:
\begin{eqnarray}\nn
    &&O_1 G_{\mu\nu}(x) = D_{\alpha_1}D_{\alpha_2}D_{\alpha_3}\tilde G_{\mu\nu}(x)\,,\\ \label{eq:operatorsOOO}
    &&O_2 G_{\mu\nu}(x) = D_{\alpha_1}D_{\alpha_2}G_{\mu\nu}(x)\,,\\\nn
    &&O_3 G_{\mu\nu}(x) = D_{\alpha_3}G_{\mu\nu}(x)\,.
\end{eqnarray}
In general, one might construct another interpolating currents which couple to $0^{--}$ state and include
four gluons~\cite{Boulanger:2008aj}, for example. However, the consideration of these states is beyond of the scope of our paper and will be the subject of our
future study.
The coefficient in the current Eq.~(\ref{eq:Jmm-general}) was chosen to have the leading term in the following form:
\begin{equation}
    J(x)\stackrel{\text{LO}}{=} g_s^3 d^{abc}  
    \tilde
    G^a_{\mu\nu;\tau_1\tau_2\tau_3}(x)~
    G^b_{\nu\ro;\tau_1\tau_2}(x)~
    G^c_{\ro\mu;\tau_3}(x)\,,
\end{equation}
where
$   G^a_{\mu\nu;\tau_1\tau_2\cdots\tau_n}=
    \partial_{\tau_1}\partial_{\tau_2}\cdots\partial_{\tau_n}G^a_{\mu\nu}
$.
Using this current, Eqs.~(\ref{eq:Jmm-general},\ref{eq:operatorsOOO}) , we have
calculated the OPE of the correlator up to the dimension-8 operators and is given by
\begin{eqnarray}\label{eq:Pi0mm:dim8}
\Pi_{\text{(OPE)}}(Q^2) &=& \Pi_{\text{(pert)}}+\Pi_{\text{(G3)}}+\Pi_{\text{(G4)}}+\cdots = \\\nn
&&
\frac{-5\alpha_s^3}{11!4\pi}Q^{20}L \\\nn
&+&
\frac{-5\pi\alpha_s^3}{3^32^5}Q^{14}
\left(\va{gG^3}-\frac{\va{J^2}}{4}\left(5+2L\right)\right)\\\nn
&+& \frac{205\pi^2\alpha_s^2}{2^{6}3^2}Q^{12}
L\va{\alpha_s^2G^4}+\cdots\,,
\end{eqnarray}
where $\alpha_s=g_s^2/(4\pi)$ is the coupling constant, $L=\ln(Q^2/\mu^2)$,  $\mu^2$ is
the  renormalization scale, the dimension-6 condensates are
$\va{gG^3}=\va{gf^{abc}G^a_{\mu\nu}G^b_{\nu\ro}G^c_{\ro\mu}}$ and
$\va{J^2}=\va{J^a_\mu J^a_\mu}$ with the quark current
$J^a_\mu=\bar q\ga_\mu t^aq$, and
the dimension-8 condensate is
\begin{eqnarray}\nn 
\va{\alpha_s^2G^4}= \va{(\alpha_s f^{abc}G^b_{\mu\nu}G^c_{\alpha\beta})^2}
-2\va{(\alpha_s f^{abc}G^b_{\mu\nu}G^c_{\nu\beta})^2}\,.
\end{eqnarray}
We adopt Mathematica package FEYNCALC~\cite{Shtabovenko:2016sxi} to handle
the algebraic manipulation.

In contrast with the previous study ~\cite{Qiao:2014vva}
mentioned above ,	
	we have a positive LO imaginary part and, therefore, we expect a consistent SR.
We would like to emphasize that the so-called direct instantons, which effect strongly the SRs for
the $0^{++}$ and $0^{-+} $ two-gluon states
~\cite{Schafer:1994fd,Harnett:2000fy,Forkel:2003mk},
 do not contribute in this case due to the symmetric color structure of the current, Eq.~(\ref{eq:Jmm-general}).

Following the method developed in ~\cite{Shifman:1978bx},
we apply the Borel transformation $\hat{B}$
\begin{eqnarray}\nn
\hat{B}_{Q^2\to M^2}\!\left[\Pi(Q^2)\right]
	= \mathop{\text{lim}}\limits_{n\to\infty}\!
	\frac{(-Q^2)^n}{\Gamma(n)}\!
	\left[\frac{d^n}{dQ^{2n}}\Pi(Q^2)\right]_{Q^2=n M^2}\
\end{eqnarray}
to both sides of the SR, Eq. (\ref{SR1}).
Using the Borel transformation allows to reduce the SR uncertainties by suppression of the contributions
from excited resonances and higher order OPE terms.
After the Borel transformation the new sum rule is
\begin{equation}
\sum_t{\cal R}^t_0(M^2,s_0)= {\cal R}^\text{(res)}_0(M^2,s_0),
\end{equation}
where $M^2$ is the Borel parameter,
\begin{eqnarray}\nn 
    {\cal R}^t_0(M^2,s_0)&=&
    \frac 1\pi \int_0^{s_0}\!\!ds~ \text{Im} \Pi_t(-s)~e^{-s/M^2}\,,\\\nn
    {\cal R}^\text{(res)}_0(M^2,s_0)&=&
    M_G^{20} F_G^2 e^{-M_G^2/M^2}\,, 
\end{eqnarray}
  and $\Pi_t$ denotes the different contributions to OPE  of the correlator:
the perturbative term (pert), and the dimension-6 (G3) and dimension-8 (G4) nonperturbative  terms.
To extract the mass from the SR, we use a family of derivative SRs obtained by differentiation
with respect to the Borel parameter $M^2$:
\begin{eqnarray}\nn 
    {\cal R}^t_k(M^2,s_0)&=& M^4\frac{d}{d M^2}{\cal R}^t_{k-1}(M^2,s_0) \,.
\end{eqnarray}

We define the difference of the OPE result and the continuum contribution as
\begin{eqnarray}\nn 
    &&{\cal R}^\text{(SR)}_k(M^2,s_0) =\\\nn
    && ~~~~~
     {\cal R}^\text{(pert)}_k(M^2,s_0)
    +{\cal R}^\text{(G3)}_k(M^2,s_0)
    +{\cal R}^\text{(G4)}_k(M^2,s_0) \,.
\end{eqnarray}
Then the master sum rule  ($k=0$) and the derivative SRs ($k>0$) can be expressed by the following equations:
\begin{eqnarray}\label{eq:RESvsSR}
    {\cal R}^\text{(SR)}_k(M^2,s_0) &\approx&  {\cal R}^\text{(res)}_k(M^2,s_0)\,.
\end{eqnarray}
The fiducial window $M^2\in [M_-^2,M_+^2]$ is limited by the conditions
that insure the reliability of the resonance model and the OPE, i.e.,
\begin{eqnarray}\label{eq:window}
&&  |{\cal R}^\text{(G4)}_k(M^2,\infty)|/{\cal R}^\text{(SR)}_k(M^2,\infty)
    <  1/3\,,\\\nn
&&  \frac{{\cal R}^\text{(res)}_k(M^2,s_0)|}{{\cal R}^\text{(SR)}_k(M^2,\infty)}\approx
    \frac{{\cal R}^\text{(SR)} _k(M^2,s_0)|}{{\cal R}^\text{(SR)}_k(M^2,\infty)}
    >  \frac 1{10}\,.
\end{eqnarray}
Then the QCD SRs for the mass and the decay constant can be presented in the form:
\begin{eqnarray}\label{eq:massSR}
M_G^k(M^2,s_0) &=&\sqrt{\frac{{\cal R}^\text{(SR)}_{k+1}(M^2,s_0)}{{\cal R}^\text{(SR)}_{k}(M^2,s_0)}}\,,\\\nonumber
F_G^k(M^2,s_0) &=&
\frac{\sqrt{e^{M_G^2/M^2}{\cal R}^\text{(SR)}_k(M^2,s_0)}}{M_G^{10}}\,.
\end{eqnarray}

\begin{figure}[h]
    \includegraphics[height=0.14\textwidth]{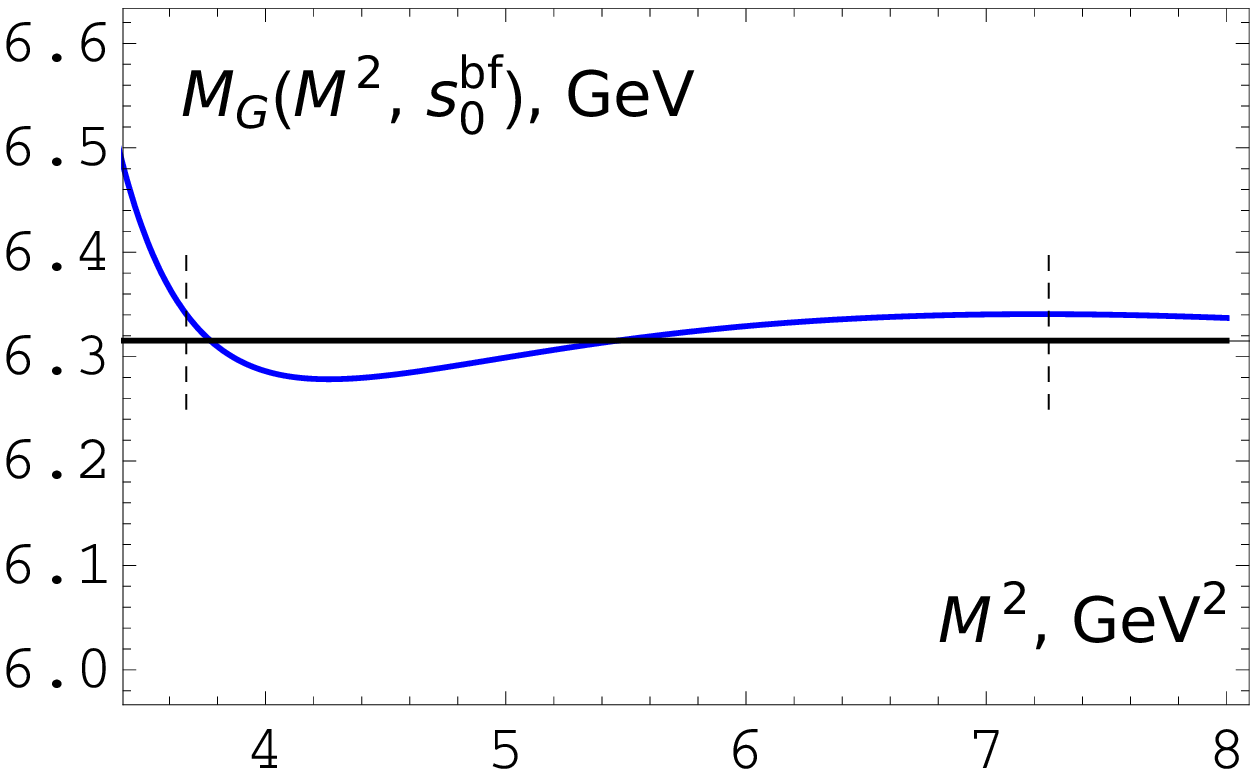}
\hfill
    \includegraphics[height=0.14\textwidth]{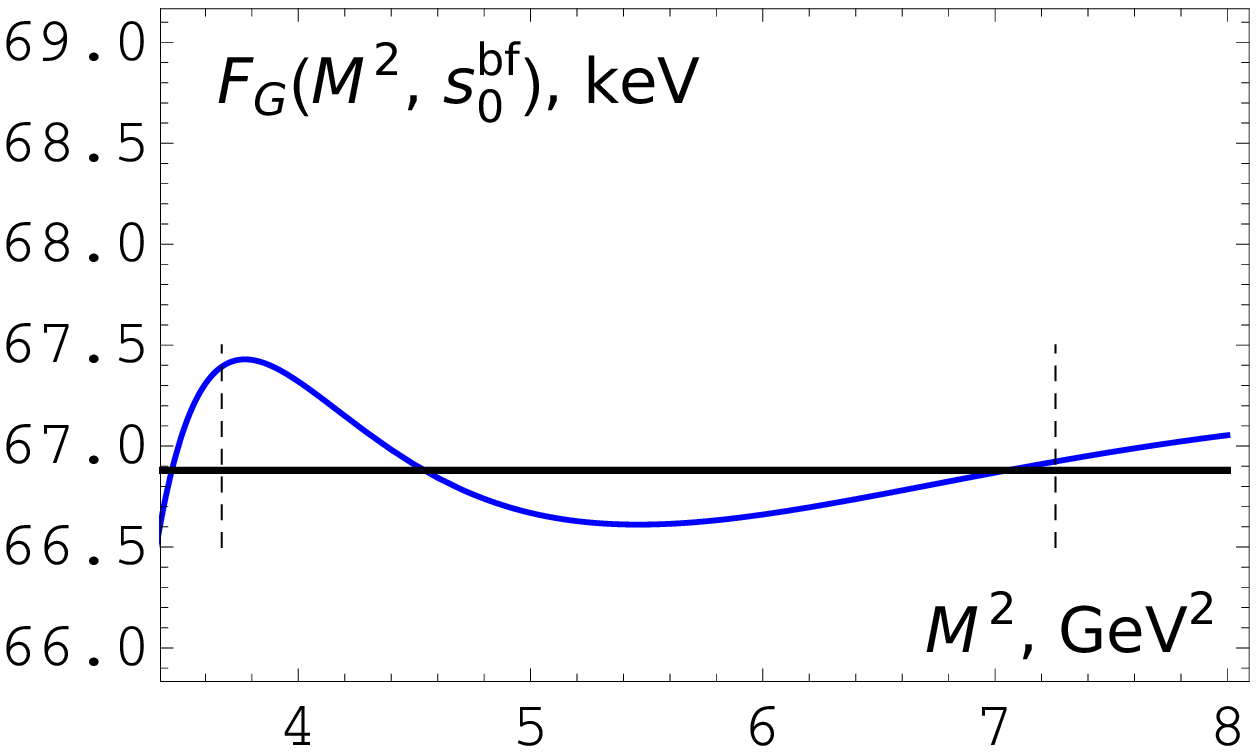}
    \caption{
        \label{fig:diagr}
        We show  the dependence on the Borel parameter of the mass (left panel) and  the decay constant (right panel)
        for the central value of the gluon condensate and best fit value of the threshold $s_0^{bf}$.
           The both panel are given for the $k=0$ case.
   The vertical lines denote the fiducial interval of the Borel parameter
   where conditions of confidence, Eq.(\ref{eq:window}), are saturated.
    The horizontal lines denote average values at fiducial interval.
}
\end{figure}

We define the mass and decay constant
by minimization of the criteria $\delta_k(s_0^\text{bf})=\delta_k^\text{min}$
with respect to the threshold $s_0$
and find the best fit value $s_0^\text{bf}$:
\begin{eqnarray}\nn
\delta_k(s_0)&=&\frac{\text{max} |M_G^k(M^2_i,s_0)-M_G^k(s_0)|}{M_G^k(s_0)}\,,\\\nn
M_G^k(s_0)&\equiv& \frac 1{n+1}\sum_{i=0}^{n} M_G^k(M^2_i,s_0)\,,
\end{eqnarray}
where we consider $n=20$ points in the fiducial interval $M^2_i=M^2_-+(M^2_+-M^2_-)~i/n$.
In Fig. \ref{fig:diagr}, we present the $k=0$ results for the glueball mass and decay constant as a function of the Borel parameter.
As one can see, we have a rather good stability plateau for both quantities.

Finally, we define the decay constant and mass as
an average in the fiducial interval for the best fit value of the threshold:
\begin{eqnarray}\nn
M_G=M_G^k(s_0^\text{bf})\,,~~
F_G^2=\frac 1{n+1}\sum_{i=1}^{n}
\frac{e^{M_G^2/M^2_i}}{M_G^{20}}
{\cal R}^\text{(SR)}_k(M^2_i,s_0^\text{bf})\,.
\end{eqnarray}

 We next follow the common practice of the renormalization group improvement of the SR:
in $\text{Im} \Pi_t(-s)$ all coupling constants are replaced by $\alpha_s\to\alpha_s(M^2)$.
 We use the strong coupling constant
\begin{eqnarray}\nn 
\alpha_s(Q^2)&=&\frac{4\pi}{b_0 \ln(Q^2/\Lambda_\text{QCD}^2)}\,,
\end{eqnarray}
with $b_0=11-2N_f/3$ and QCD scale $\Lambda_\text{QCD}=300$~MeV.
 Since we are working in gluodynamics, we put the number of flavors $N_f=0$ and eliminate the quark and quark-gluon condensate contributions.
The dimension-6 three-gluon condensate $ \va{gG^3}$ doesn't contribute here
due to absence of the correspondent  $\ln(Q^2)$ terms in the correlator, Eq. (\ref{eq:Pi0mm:dim8}).
For the dimension-8 gluon condensate the hypothesis of vacuum dominance yields the relation
\begin{equation}
\va{\alpha_s^2G^4}=\frac{3}{2^4}\va{\alpha_sG^2}^2.
\nn
\end{equation}

In our case, the mass of the exotic glueball is determined by the squared value of the gluon condensate
$\va{\alpha_sG^2}=\va{\alpha_sG^a_{\mu\nu}G^a_{\mu\nu}}$.
Unfortunately, this value is not well known.
Following the analyses carried out in refs.\cite{Vainshtein:1978wd,Novikov:1983jt,Ioffe:2005ym,Narison:2002pw},
we take
\begin{equation}\nn 
\va{\frac{\alpha_s}{\pi}G^2}=0.012\pm 0.006~\text{GeV}^4.
\end{equation}
Implementing the QCD SR analysis described above we obtain for the prediction of
the mass and the decay constant
 from the  $k=0$ SR (see Eqs.(\ref{eq:RESvsSR},\ref{eq:massSR}))
\begin{eqnarray}\label{mass}
M_G&=&6.3^{+0.8}_{-1.1}~\text{GeV}\,,~~~
F_G=67 \pm 6 ~\text{keV}\,.
\end{eqnarray}

The mass and decay constant estimates for the higher values of $k=1,2,3$ are in agreement,
within the error bars,  with $k=0$ case considered.
The SR analysis in full QCD (number of flavors $N_f=3$ and nonzero quark condensate $\va{J^2}$)
leads to a reduction of the glueball mass by $0.2$~GeV.
The  mass of the exotic  glueball in Eq. (\ref{mass}) is not far away  from
the recent unquenched lattice result $M_G=5.166\pm 1.0$ GeV~\cite{Gregory:2012hu}
obtained with a rather large pion mass $m_\pi=360$ MeV.

Here we would like to note that there are three sources of uncertainties in the above analysis for the mass and decay constant:
i) the variation of the gluon condensate; ii) the stability of the SR triggering
the Borel parameter $M^2$ dependence in terms of the criteria $\delta_k^\text{min}$;
and iii) the roughly estimated SR uncertainty coming from the  OPE truncation.
The latter  uncertainty for the  decay constant comes from the definition of the fiducial interval, Eq.~(\ref{eq:window}),
in the standard assumption that the contribution from the missing terms is of the order of
the last included nonperturbative term squared: $(1/3)^2\sim10\%$.
The same error for the mass can be expected to be suppressed
since the related errors for
${\cal R}^\text{(SR)}_{k+1}$ and
${\cal R}^\text{(SR)}_{k}$ are correlated.
The presumable underestimation of uncertainties related to the OPE truncation is
unlikely due to conservative choice of the gluon condensate uncertainty.
The considered three sources of uncertainty can be given in percentage of the final uncertainty
for the mass and  the decay constant
\begin{eqnarray}\nn
M_G&=&6.3^{+12\%}_{-17\%}\pm 0.5\% \pm 0\%  ~\text{GeV}\,,
\\\nn
F_G&=&67^{+2\%}_{-3\%} \pm 0.6\%  \pm 5\% ~\text{keV}\,.
\end{eqnarray}
where
the first uncertainty is related to gluon condensate variation,
the second is representing the stability of SR, and
the third  is OPE truncation uncertainty.

The best fit threshold value is $s_0^\text{bf}=52.4^{+12.6\%}_{-16.2\%}$~GeV$^2$
when only  the uncertainty of the gluon condensate is included. Note that the fiducial interval for the  central value of the gluon condensate
is $M^2\in [3.7, 7.3]$~GeV$^2$.

The glueball width can be estimated in the QCD SR approach also using the broad resonance distribution.
The good stability of the zero width resonance based SR, Eq. (\ref{SR1}), shows that
we can extract only the upper limit of the glueball width from the QCD SR.
The simplest way to introduce the width is by using unit step functions~\cite{Harnett:2000fy}:
\begin{eqnarray}\nn
&&\text{Im}\Pi^{\text{(res2)}}(-s)=\\\nn
&&\frac{\pi (m^2)^{N-2}f^2}{2m\Gamma} \left(\Theta(s-m^2+m\Gamma)-\Theta(s-m^2-m\Gamma)\right)\,.
\end{eqnarray}
Requiring that the stability of the broad resonance based SR is better than
the stability of zero width resonance based SR,
\begin{equation}
\max\left|1-\frac{{\cal R}^\text{(res2)}_k(M^2_i,s_0)}{{\cal R}^\text{(SR)}_k(M^2_i,s_0)}\right|
\leq
\max\left|1-\frac{{\cal R}^\text{(res)}_k(M^2_i,s_0)}{{\cal R}^\text{(SR)}_k(M^2_i,s_0)}\right|\,,
 \nn
\end{equation}
we obtain an upper limit for the glueball width, $\Gamma_G\leq 235$~MeV.
The used stability test was chosen for the simplicity and transparency of the width estimation
keeping the level of accuracy at the level of SR accuracy for mass and decay constant.
In the new SR we vary only the width value while the values for condensate,
mass and  decay constant remain fixed.
The Borel parameter value is varied in the interval $M^2_i\in [3.7, 7.3]$~GeV$^2$.
This result indicates  that the  $0^{--}$ glueball should be rather narrow. Therefore, it can be seen in the appropriate
experiments.

By quantum numbers the exotic glueball could mix with the exotic $0^{--}$ tetraquark.
However, a recent study with QCD SR for this tetraquark has obtained a small mass,
$M_{tetra}=1.66\pm 0.14 $ GeV,\cite{Huang:2016rro}.
The large mass difference between the two states leads us to expect a very small mixing between them.
Thus, we can consider the exotic $0^{--}$ glueball as a pure gluon state.

Summarizing, we have presented a  QCD SR study  for the exotic three-gluon glueball
state with quantum numbers $J^{PC}=0^{--}$ using a new interpolating current.
We have analyzed the QCD SR consisting of contributions of operators up to dimension-8 and have obtained an estimation
 of the mass, the decay constant and an upper limit for the width of  the  exotic glueball.
 These results provide a clear guide for the search of this important state in the experiments.

 After the paper was completed we were informed of the negative result of  the search of the low mass
 exotic $0^{--}$ glueball by the Belle Collaboration \cite{Jia:2016cgl}.

We would like to thank J. Evslin, B.~Gudnason, M.~Ivanov, S. Mikhailov, and, especially,  V. Vento
 for stimulating discussions and useful remarks.
This work has been supported by the National Natural Science Foundation of China (Grants No. 11575254 and 11650110431),
  Chinese Academy of Sciences President's International Fellowship Initiative
  (Grant No. 2013T2J0011 and 2016PM053), the Japan Society for the Promotion
of Science (Grant No.S16019).
The work by H.J.L. was supported by the Basic Science Research Program
through the National Research Foundation of Korea (NRF) funded by Ministry
of Education under Grants No. 2013R1A1A2009695.
This work was also supported by the Heisenberg--Landau Program (Grant 2016),
the Russian Foundation for Basic Research under Grants No.\ 15-52-04023.

\end{document}